\newcommand{\beq}{\begin{equation}}
\newcommand{\eeq}{\end{equation}}

\newcommand{\id}
 {i\kern.06em\hbox{\raise.25ex\hbox{$/$}\kern-.60em$\partial$}}

\newcommand{\bs}{/\kern-.52em b}
\newcommand{\ds}{/\kern-.52em d}
\newcommand{\qs}{/\kern-.52em s}
\newcommand{\cD}{{\cal D}}

\newcommand{\p}{\partial}
\newcommand{\yp}{^{\prime}}

\newcommand{\dd}
{\kern.06em\hbox{\raise.25ex\hbox{$/$}\kern-.60em$\partial$}}

\newcommand{\ep}{\epsilon}

\documentstyle[12pt]{article}
\date{}
\begin{document}
\title{Conformal Field Theory Correlators From sine-Gordon Model on AdS Spacetime \footnotetext{\# e-mail:sshfeng@yahoo.com; ssfeng@ustc.edu.cn}}
\author{{Sze-Shiang Feng $^{1,2,\#}$, Mu-Lin Yan $^2$}\\
1.{\small {\it CCAST(World Lab.), P.O. Box 8730, Beijing 100080}}\\
2. {\small {\it Department of Astronomy and Applied Physics}}\\
{\small {\it University of Science and Technology of China, 230026, Hefei, China}}}
\maketitle
\newfont{\Bbb}{msbm10 scaled\magstephalf}
\newfont{\frak}{eufm10 scaled\magstephalf}
\newfont{\sfr}{eufm7 scaled\magstephalf}
\baselineskip 0.2in
\begin{center}
\begin{minipage}{135mm}
\vskip 0.3in
\baselineskip 0.2in
\begin{center}{\bf Abstract}\end{center}
  {Using the proposed AdS/CFT correspondence, we calculate the correlators of operators of conformal field theory at the boundary of AdS$_{d+1}$ corresponding to the sine-Gordon model in the bulk.
  \\PACS number(s):11.25.Hf,11.10.Kk
   \\Key words: conformal field theory, sine-Gordon, AdS spacetime }
\end{minipage}
\end{center}
\vskip 1in
\section{Introduction}
The well-known Maldacena conjecture proposed two years ago\cite{s1}\cite{s2} brought to the string theory a revolution: it unravels a relation between Yang-Mills theory and string theory and thus sheds light on the final unification of all interactions. Although the original conjecture referred to the type-IIB string theory on AdS$_5\times S^5$ and the ${\cal N}=4, d=3+1 $U(N) super-Yang-Mills theory, it seems valid in general: a quantum field theory in the bulk of d+1 dimensional anti-de Sitter spacetime (AdS$_{d+1}$) with some boundary conditions corresponds to a conformal field theory(CFT) on the boundary. Though an exact proof of the conjecture is still lacking, its validity can be confirmed by a number of tests\cite{s3}-\cite{s11}. 
The mathematical scheme of the correspondence was formulated explicitly by Witten
\cite{s4} and independently by Polyakov {\it et al} \cite{s5}. For a scalar field $\phi$ in AdS$_{d+1}$, it is
\beq
Z_{{\rm AdS}}[\phi_0]=\int_{\phi_0}\cD\phi\exp(-I[\phi])\equiv Z_{{\rm CFT}}[\phi_0]=<\exp(\int_{\p\Omega}d^dx {\cal O}\phi_0)>
\eeq
The path-integral on the l.h.s. is calculated under the restriction that the field $\phi$ approaches to $\phi_0$ on the boundary. The correspondence says that this path-integral is to be identified with the r.h.s. which is the partition of a CFT on the boundary with $\phi_0$ playing the role of a current coupled to a conformal operator ${\cal O}$. The r.h.s. thus enables us to calculate the correlators of ${\cal O}$ of the CFT on the boundary. Since the conformal invariance determines the 2- and 3-point functions, the nontrivial ones are the cases for $n\ge 3$. (
 So far, the correspondence (1) is still a kind of guesswork which is only more explicit than the original Maldacena's statement. There must be some field-theoretic reason behind\cite{s12}. One may regard (1) as the dilaton sector of the type-IIB string theory and corresponding CFT is the corresponding sector of the full super-Yang-Mills theory.)\\
\indent The free $\phi$-theory was considered in \cite{s3} and the general interacting cases were studied in \cite{s5}. Yet a careful investigation shows that the consideration of \cite{s5} is not enough for the case of the sine-Gordon(sG) theory. This is the very purpose of the present paper. 
The action of the sG theory in AdS spacetime is
\beq
I[\phi]=\int_{\Omega}d^{d+1}x\sqrt{g}[\frac{1}{2}(\nabla\phi)^2-\frac{m^2}{\beta^2}(\cos\beta\phi-1)]
\eeq 
The classical equation of motion reads
\beq
\nabla^2\phi-\frac{m^2}{\beta}\sin\beta\phi=0
\eeq
or
\beq
\nabla^2\phi-m^2\phi=m^2(\frac{1}{\beta}\sin\beta\phi-\phi)\equiv J(\phi)
\eeq
The free field case is a limit of $\beta\rightarrow 0$. If we write
\beq
\frac{m^2}{\beta^2}(1-\cos\beta\phi)=\frac{1}{2}m^2\phi^2+\sum_{n\ge 3}\frac{\lambda_n}{n!}\phi^n
\eeq
we have
\beq
\lambda_{2n+1}=0, \,\,\,\,\,\,\, \lambda_{2n}=(-1)^{2n}m^2\beta^{2(n-1)}
\eeq
Eq(4) can also be cast into the form
\beq
(\nabla^2-m^2)\phi=\sum_{n\ge 3}\frac{\lambda_n}{(n-1)!}\phi^{n-1}
\eeq
as the eq(3) in \cite{s5}. The difference is that here there is only one parameter:$\beta$. As usual, the dominant contribution to the path-integral in the l.h.s. of (1) comes from the classical path satisfying the equation of motion (3). So as an approximation, we are for the moment interested in the classical solution to (3) with a given Dirichlet boundary condition $\phi(x_0,{\bf x})_{\mid\p\Omega}=\phi_0$. We would like to emphasize that the nonlinearity of the equation of motion renders it to have possibly more than one solutions to the Dirichlet problem. It is a nontrivial problem whether different solutions lead to the same correlators of the boundary CFT. Using the covariant Green's function satisfying
\beq
(\nabla^2-m^2)G(x,y)=\frac{\delta(x-y)}{\sqrt{g(x)}}
\eeq
and the boundary condition
$G(x,y)_{\mid x\in\p\Omega}=0$. The classical equation of motion can be expressed equivalently as
\beq
\phi(x)=\int_{\p\Omega}d^d{\bf y}\sqrt{h}n^{\mu}\frac{\p}{\p y^\mu}G(x,y)\phi(y)
+\int_{\Omega}d^{d+1}y\sqrt{g(y)}G(x,y)\sum_{n\ge 3}\frac{\lambda_n}{(n-1)!}\phi^{n-1}
\eeq
This integral equation can be employed to obtain approximate solutions by recursion.\\
\indent  In section 2 we give a review of the study of free $\phi$ case while presenting a detailed derivation of the connection of the result in\cite{s3} and that in \cite{s5} of the $\ep$-boundary problem. In section 3 we study the case for sG theory.
\section{The free-$\phi$ case: a review}
We use the representation of AdS$_{d+1}$ as the upper half-space($x_0>0$) with the metric
\beq
ds^2=\frac{1}{x_0^2}\sum^d_{i=0}dx_i^2
\eeq
The scalar curvature $R=-d(d+1)$. The boundary of AdS is identified with $x=0$ and the single point $x_0=\infty$. The solution of the classical equation of motion
\beq
(\nabla^2-m^2)\phi=[x_0^2\sum^d_{i=0}\p^2_i-x_0(d-1)\p_0-m^2]\phi=0
\eeq
can be obtained
\beq
x_0^{d/2}e^{-i{\bf k}\cdot{\bf x}}I_{\alpha}(kx_0);\,\,\,\, x_0^{d/2}e^{-i{\bf k}\cdot{\bf x}}K_{\alpha}(kx_0) 
\eeq
where $\alpha=\sqrt{\frac{d^2}{4}+m^2}$. ${\bf k}$ is the momentum $d$-vector and $k=|{\bf k}|$. $ I_{\alpha} $ and $K_{\alpha}$ are the Bessel functions. The modes in (12 are linearly independent and constitute a complete basis of the Hilbert space. As in quantum mechanics, the Green's function can be expressed as $G(x,y)=\sum_n\frac{\psi_n(x)\psi^{\star}_n(y)}{\lambda_n}, \lambda_n$ are the eigenvalues corresponding to the eigenfunctions $\psi_n$, here we have
\beq
G(x,y)=\int\frac{d^d{\bf k}}{(2\pi)^d}x_0^{d/2}e^{-i{\bf k}\cdot({\bf x}-{\bf y})}(-y_0^{d/2})
\{
\matrix{I_{\alpha}(kx_0)K_{\alpha}(ky_0) & {\rm for}\,\,\,\, x_0<y_0\cr
K_{\alpha}(kx_0)I_{\alpha}(ky_0) & {\rm for} \,\,\,\,x_0>y_0\cr}
\eeq
It can also be expressed as\cite{s5}
\beq
G(x,y)=-\frac{c}{2\alpha}\xi^{-\Delta}F(\frac{d}{2},\Delta;\alpha+1,\frac{1}{\xi^2})
\eeq
where $F$ denotes the hypergeometric function and
\beq
\xi=\frac{1}{2x_0y_0}[\frac{1}{2}((x-y)^2+(x-y^*)^2)+\sqrt{(x-y)^2(x-y^*)^2}]
\eeq
in which $y^*=(-y_0,y)$. $\Delta=d/2+\alpha,c=\Gamma(\Delta)(\pi^{d/2}\Gamma(\alpha))$.\\
\indent Since the classical solution of $\phi$ in terms of the boundary value
$\phi_0$ involves the determinant of metric at the boundary which is singular, the $\epsilon$-description of the asymptotic boundary is necessary: one first solve the problem at $x_0=\ep$ and then take the limit $\ep\rightarrow 0$ in the very end. The corresponding Green's function is
\beq
G_{\ep}(x,y)=G_0(x,y)+\int \frac{d^d{\bf k}}{(2\pi)^d}(x_0y_0)^{d/2} e^{-i{\bf k}\cdot({\bf x}-{\bf y})}K_{\alpha}(kx_0)K_{\alpha}(ky_0)\frac{I_{\alpha}(k\ep)}{K_{\alpha}(k\ep)}
\eeq
Now we calculate the normal derivative of $G_{\ep}$ at the $\ep$-boundary. For $x_0\ge y_0$ since
\beq
G_{\ep}(x,y)=\int\frac{d^d{\bf k}}{(2\pi)^d} e^{-i{\bf k}\cdot({\bf x}-{\bf y})}x_0^{d/2}K_{\alpha}(kx_0)y_0^{d/2}[K_{\alpha}(ky_0)\frac{I_{\alpha}(k\ep)}{K_{\alpha}(k\ep)}-I_{\alpha}(ky_0)]
\eeq
so
\beq
\frac{\p}{\p y_0}G_{\ep}(x,y)_{\mid y_0=\ep}=\int\frac{d^d{\bf k}}{(2\pi)^d} e^{-i{\bf k}\cdot({\bf x}-{\bf y})}x_0^{d/2}K_{\alpha}(kx_0)\ep^{d/2}\frac{1}{K_{\alpha}(k\ep)}(kK_{\alpha}^{\yp}(ky_0)I_{\alpha}(k\ep)- kK_{\alpha}(k\ep)I^{\yp}_{\alpha}(ky_0))_{\mid y_0=\ep}
\eeq
Using
\beq
K_{\alpha}(z)=\frac{\pi}{2\sin\alpha\pi}[I_{-\alpha}(z)-I_{\alpha}(z)]
\eeq
we have the Wronskian determinant
\beq
\left|\matrix{I_{\alpha} & K_{\alpha}\cr I^{\yp}_{\alpha} & K^{\yp}_{\alpha}\cr }\right|=\frac{\pi}{2\sin\alpha\pi}\left|\matrix{I_{\alpha} & I_{-\alpha}\cr I^{\yp}_{\alpha} & I^{\yp}_{-\alpha}\cr }\right|
\eeq
Since $I_{\alpha}(z)=e^{-\frac{\nu\pi}{2}i}J_{\alpha}(iz)$ we have
\beq
\left|\matrix{I_{\alpha} & I_{-\alpha}\cr I^{\yp}_{\alpha} & I^{\yp}_{-\alpha}\cr }\right|=-\frac{2\sin\alpha}{\pi z}
\eeq
therefore
\beq
(kK_{\alpha}^{\yp}(ky_0)I_{\alpha}(k\ep)- kK_{\alpha}(k\ep)I^{\yp}_{\alpha}(ky_0))_{\mid y_0=\ep}=-\ep^{-1}
\eeq
Thus
\beq
\frac{\p}{\p y_0}G_{\ep}(x,y)_{\mid y_0=\ep}=-x_0^{d/2}\ep^{d/2-1}\int\frac{d^d{\bf k}}{(2\pi)^d} e^{-i{\bf k}\cdot({\bf x}-{\bf y})}\frac{K_{\alpha}(kx_0)}{ K_{\alpha}(k\ep)}
\eeq
Now we study the asymptotic behavior of the l.h.s. of (23). Using
\beq
\lim_{z\rightarrow 0}z^{\alpha}K_{\alpha}(z)=2^{\alpha-1}\Gamma({\alpha})
\eeq
we have in the limit $\lim_{z\rightarrow 0}$
\beq
\int\frac{d^d{\bf k}}{(2\pi)^d} e^{-i{\bf k}\cdot({\bf x}-{\bf y})}\frac{K_{\alpha}(kx_0)}{ K_{\alpha}(k\ep)}
=\frac{\ep^{\alpha}}{2^{\alpha-1}\Gamma(\alpha)} \int\frac{d^d{\bf k}}{(2\pi)^d} e^{-i{\bf k}\cdot {\bf r}}k^{\alpha}K_{\alpha}(kx_0)
\eeq
where ${\bf r}={\bf x}-{\bf y}$. We write
\beq
\int\frac{d^d{\bf k}}{(2\pi)^d} e^{-i{\bf k}\cdot {\bf r}}k^{\alpha}K_{\alpha}(kx_0)=2\int^{+\infty}_{-\infty}dt\int^{+\infty}_0\frac{dk_0}{2\pi}k_0^{\alpha+1}K_{\alpha}(k_0x_0)e^{itk_0^2}\int\frac{d^d{\bf k}}{(2\pi)^d}e^{-it{\bf k}^2-i{\bf k}\cdot{\bf r}}
\eeq
Using
\beq
\int\frac{d^d{\bf k}}{(2\pi)^d}e^{-it{\bf k}^2-i{\bf k}\cdot{\bf r}}=e^{i\frac{{\bf r}^2}{4t}}\frac{\pi^{d/2}}{(2\pi)^d\mid 
t\mid^{d/2}}\{\matrix{e^{i\pi d/4}\,\,\,\, t<0\cr e^{-i\pi d/4}\,\,\,\, t>o}
\eeq
We have
\beq
\int\frac{d^d{\bf k}}{(2\pi)^d} e^{-i{\bf k}\cdot {\bf r}}k^{\alpha}K_{\alpha}(kx_0)=\frac{2\pi^{d/2}}{(2\pi)^d}\int^{+\infty}_0dt\int^{+\infty}_0\frac{dk_0}{2\pi}k_0^{\alpha+1}K_{\alpha}(k_0x_0)t^{-d/2}[ e^{-i\pi d/4}e^{i(tk_0^2+\frac{{\bf r}^2}{4t})}+c.c]
\eeq
Using (formula 3.471-11 in \cite{s13})
\beq
\int^{+\infty}_0x^{\nu-1}e^{i\frac{\mu}{2}(x+\beta^2/x)}dx=i\pi\beta^{\nu}e^{-i\nu\pi/2}H^{(1)}_{-\nu}(\beta\mu)
\eeq
($H^{(1)}_{-\nu}(\beta\mu)$ denotes the Hankel functions),we have
\beq
\int^{+\infty}_0dt t^{-d/2}e^{i(tk_0+\frac{{\bf r}^2}{4t})}=i\pi(\frac{r}{2k_0})^{\nu}e^{-i\nu\pi/2}H^{(1)}_{d/2-1}(k_0r)=
(\int^{+\infty}_0dt t^{-d/2}e^{-i(tk_0+\frac{{\bf r}^2}{4t})})^*
\eeq
Therefore the integral
$$
\int^{+\infty}_0dk_0k_0^{\alpha+d/2}K_{\alpha}(k_0x_0)H^{(1)}_{d/2-1}(k_0r)$$
\beq
=
\int^{+\infty}_0dk_0k_0^{\Delta}K_{\alpha}(k_0x_0)\frac{i}{\sin(\frac{d}{2}-1)\pi}[e^{-i\pi(d/2-1)}J_{d/2-1}(k_0r)-J_{1-d/2}(k_0r)]
\eeq
is involved. Using (formula 6.576-1 in \cite{s13})
\beq
\int^{+\infty}_0dx x^{-\lambda}K_{\mu}(ax)J_{\nu}(bx)=\frac{b^{\nu}\Gamma(\frac{\nu-\lambda+\mu+1}{2})\Gamma(\frac{\nu-\lambda-\mu+1}{2})}{2^{\lambda+1}a^{\nu-\lambda+1}\Gamma(1+\nu)}F(\frac{\nu-\lambda+\mu+1}{2},\frac{\nu-\lambda-\mu+1}{2};\nu+1;-\frac{b^2}{a^2})
\eeq
we have
\beq
\int^{+\infty}_0\frac{dk_0}{2\pi} k_0^{\Delta}K_{\alpha}(k_0x_0)J_{d/2-1}(k_0r)=\frac{r^{d/2-1}\Gamma(\Delta)}{2\pi 2^{1-\Delta}x_0^{d/2+\Delta}}(1+\frac{r^2}{x_0^2})^{-\Delta}
\eeq
where we have used the formula $F(-\alpha,\beta;\beta;-z)=(1+z)^{\alpha}$. Similarly
\beq
\int^{+\infty}_0\frac{dk_0}{2\pi} k_0^{\Delta}K_{\alpha}(k_0x_0)J_{d/2-1}(k_0r)=0
\eeq
where we have used that $ \Gamma(1)=0$. Therefore
\beq
\int^{+\infty}_0\frac{dk_0}{2\pi}k_0^{\alpha+d/2}K_{\alpha}(k_0x_0)H^{(1)}_{d/2-1}(k_0r)=
\frac{i}{\sin(d/2-1)\pi}e^{-i\pi(d/2-1)}\frac{r^{d/2-1}\Gamma(\Delta)}{2\pi 2^{1-\Delta}x_0^{d/2+\Delta}}(1+\frac{r^2}{x_0^2})^{-\Delta}
\eeq
Note that $K_{\nu}$ is real, we have hence
\beq
\int\frac{d^d{\bf k}}{(2\pi)^d} e^{-i{\bf k}\cdot {\bf r}}k^{\alpha}K_{\alpha}(kx_0)=
\frac{2^{d/2+\Delta}\pi^{d/2+1}\Gamma(\Delta)}{(2\pi)^{d+1}}
\frac{x_0^{\Delta-d/2}}{(x_0^2+r^2)^{\Delta}}
\eeq
Therefore in the limit $\lim_{\ep\rightarrow 0}$
\beq
\frac{\p}{\p y_0}G_{\ep}(x,y)_{\mid y_0=\ep}=-\ep^{\Delta-1}c(\frac{x_0}{x_0^2+r^2})^{\Delta}
\eeq
We thus have the solution to the $\ep$-boundary problem
\beq
\phi(x)=c\ep^{\Delta-d}\int d^d{\bf y}\phi_{\ep}({\bf y})(\frac{x_0}{x_0^2+\mid{\bf x}-{\bf y}\mid^2})^{\Delta}
\eeq
Defining the boundary value of $\phi$ at $\ep=0$ as
\beq
\phi_0({\bf x})=\ep^{\Delta-d}\phi_{\ep}({\bf x})
\eeq
we can arrive at the result of \cite{s3} obtained in a tricky way. \\
\indent Using the expression for the $\ep$-boundary problem
\beq
\p_0\phi_{\mid x_0=\ep}=2\alpha c\ep^{2\alpha-1}\int d^d{\bf y}\frac{\phi_{\ep}}{\mid{\bf x}-{\bf y}\mid^{2\Delta}}+...
\eeq
we find the value of the free action as
\beq
I_{\rm free}=-\frac{1}{2}\int d^d{\bf x}d^d{\bf y}2\alpha c\ep^{2(\Delta-d)}
\frac{\phi_\ep({\bf x})\phi_{\ep}({\bf y})}{\mid{\bf x}-{\bf y}\mid^{2\Delta}}+...
\eeq
Obviously the limit $\lim_{\ep\rightarrow 0}$ makes sense. We can readily get the 2-point correlator of the CFT on the boundary
\beq
<{\cal O}({\bf x}){\cal O}({\bf y})>=\frac{2\alpha c}{\mid{\bf x}-{\bf y}\mid^{2\Delta}}
\eeq
\section{The sine-Gordon theory}
Equation (9) can be written as
\beq
\phi(x)=\int_{\p\Omega}d^d{\bf y}K(x_0,{\bf x};{\bf y})\phi({\bf y})+\int_{\Omega}
\sqrt{g(y)}G(x,y)J[\phi(y)]
\eeq
where
\beq
K(x_0,{\bf x};{\bf y})=c\frac{x_0^d}{(x_0^2+{\bf x}^2)^d}
\eeq
We search for a series solution in terms of powers of $\beta$ for the sG theory. Defining $\varphi_n$ by
\beq
\varphi_0(x)=\int_{\p\Omega} d^d{\bf y}K(x_0,{\bf x};{\bf y})\phi_0({\bf y})
\eeq
\beq
\varphi_1(x)=\int_{\Omega}d^{d+1}y\sqrt{g(y)}G(x,y)J[\varphi_0(y)]
\eeq
\beq
\varphi_2(x)=\int_{\Omega}d^{d+1}y\sqrt{g(y)}G(x,y)J[\varphi_0(y)+\varphi_1(y)]
\eeq
{\it et al}.
then
\beq
\varphi(x)=\varphi_0+\varphi_1+\varphi_2+...
\eeq
We assume that this series converges. As an approximation, we take the first two terms
\beq
\phi(x)\approx\varphi_0+\varphi_1
\eeq
Then the action corresponding to this classical path is
\begin{eqnarray}
I[\phi]&=&\int_{\Omega}d^{d+1}x\sqrt{g(x)}\{\frac{1}{2}\nabla^{\mu}\varphi_0\nabla_{\mu}\varphi_0+\nabla^{\mu}\varphi_0\nabla_{\mu}\varphi_1+\frac{1}{2}\nabla^{\mu}\varphi_1\nabla_{\mu}\varphi_1+\frac{m^2}{\beta^2}[1-\cos\beta(\varphi_0+\varphi_1)]\}\\
&=&\int_{\Omega}d^{d+1}x\sqrt{g(x)}\{\frac{1}{2}[\nabla^{\mu}\varphi_0\nabla_{\mu}\varphi_0+m^2(\varphi_0+\varphi_1)^2]+\nabla^{\mu}\varphi_0\nabla_{\mu}\varphi_1+\frac{1}{2}\nabla^{\mu}\varphi_1\nabla_{\mu}\varphi_1\\
& &
+\frac{m^2}{\beta^2}[1-\cos\beta(\varphi_0+\varphi_1)]-\frac{m^2}{2}(\varphi_0+\varphi_1)^2\}
\end{eqnarray}
We consider the second term
\beq
\int_{\Omega}d^{d+1}x\sqrt{g(x)}\nabla^{\mu}\varphi_0\nabla_{\mu}\varphi_1
=\int_{\p\Omega_{\ep}}\sqrt{h}d^d{\bf x}n^{\mu}\varphi_1\p_{\mu}\varphi_0
\eeq
Since $G(x,y)_{\mid\p\Omega_{\ep}}=0$, we see that $\varphi_{1\mid\p\Omega_{\ep}}=0$. Therefore
\beq
I[\phi]=I^{(0)}[\phi]+ \int_{\Omega}d^{d+1}x\sqrt{g(x)}\{m^2\varphi_0\varphi_1+\frac {1}{2}m^2\varphi_1^2+\frac{1}{2}\nabla^{\mu}\varphi_1\nabla_{\mu}\varphi_1
+\frac{m^2}{\beta^2}[1-\cos\beta(\varphi_0+\varphi_1)]-\frac{m^2}{2}(\varphi_0+\varphi_1)^2\}
\eeq
where
\beq
I^{(0)}[\phi]= \int_{\Omega}d^{d+1}x\sqrt{g(x)}\frac{1}{2}(\nabla^{\mu}\varphi_0\nabla_{\mu}\varphi_0+m^2\varphi_0^2)
\eeq
We expand the rest part of $I[\phi]$ in terms of the powers of $\beta$. Since
\beq
\varphi_1(x)=\int_{\Omega}d^{d+1}y\sqrt{g(y)}G(x,y)\sum_{n=1}^{\infty}
\frac{(-1)^nm^2\beta^{2n}}{(2n+1)!}\varphi_0(y)
\eeq
we see that $\varphi_1\sim\beta^2, \varphi_0\varphi_1\sim\beta^2, \varphi_1^2\sim\beta^4.$ So the part of order $O(\beta^2)$ of $I[\phi]$ is
\begin{eqnarray}
I^{(2)}[\phi]&=&\int_{\Omega}d^{d+1}x\sqrt{g(x)}[m^2\varphi_0(x)\int_{\Omega} d^{d+1}y\sqrt{g(y)}G(x,y)\frac{(-m^2)\beta^2}{3!}\varphi_0^3(y)+\frac{1}{4!}
m^2\beta^2\varphi_0^4]\\
&=&\int_{\p\Omega}d^d{\bf x}_1d^d{\bf x}_2d^d{\bf x}_3d^d{\bf x}_4
\phi_0({\bf x}_1) \phi_0({\bf x}_2)\phi_0({\bf x}_3)\phi_0({\bf x}_4)\\
& &[\frac{c^4m^2\beta^2}{4!}I_4({\bf x}_1,{\bf x}_1,{\bf x}_3,{\bf x}_4)-
\frac{m^4\beta^2}{3!}J_4({\bf x}_1,{\bf x}_1,{\bf x}_3,{\bf x}_4)]
\end{eqnarray}
where
\beq
I_4({\bf x}_1,{\bf x}_1,{\bf x}_3,{\bf x}_4)=\int_{\Omega}d^{d+1}y\frac{y_0^{4\Delta-(d+1)}}
{[(y_0^2+\mid{\bf y}-{\bf x}_1\mid^2)(y_0^2+\mid{\bf y}-{\bf x}_2\mid^2) (y_0^2+\mid{\bf y}-{\bf x}_3\mid^2) (y_0^2+\mid{\bf y}-{\bf x}_4\mid^2)]^{\Delta}}
\eeq
introduced in \cite{s5} and
\beq
J_4({\bf x}_1,{\bf x}_1,{\bf x}_3,{\bf x}_4):=\int d^{d+1}xd^{d+1}y(x_0y_0)^{-(1+d)}G(x,y)K(x_0,{\bf x};{\bf x}_1) K(y_0,{\bf y};{\bf x}_2) K(y_0,{\bf y};{\bf x}_3) K(y_0,{\bf y};{\bf x}_4)
\eeq
So we have
\beq
<{\cal O}({\bf x}_1) {\cal O}({\bf x}_2)
{\cal O}({\bf x}_3)
{\cal O}({\bf x}_4)>=
-c^4m^2\beta^2 I_4({\bf x}_1,{\bf x}_1,{\bf x}_3,{\bf x}_4)+
4m^4\beta^2 J_4({\bf x}_1,{\bf x}_1,{\bf x}_3,{\bf x}_4)
\eeq
It is seen that the both terms in (62) are of the same order so are equally important. It is a pity that both functions $I_4$ and $J_4$ are not  analytically integrable.
The two-point correlator is the same as that for the free $\phi$ theory and the 3-point correlator vanishes for the sG theory.
\section{Discussions}
   In this paper we studied the correlators of CFT on the boundary of AdS$_{d+1}$ corresponding to the sine-Gordon theory in the bulk by the AdS/CFT correspondence.
It is found that apart from $I_4$ in \cite{s5}, there is another term which makes a contribution of the same order. So the general consideration for interacting $\phi$-theories in \cite{s5} is not enough and a more careful consideration is necessary for different specific theories. We stress that in general, the interacting $\phi$ theory may have more than one solutions to the classical Dirichlet problem and it is to be answered that they lead to the same correlators for the boundary CFT. It seems that the answer is yes, since perturbatively, the answer is pisitive and the AdS/CFT correspondence seems to ensure the uniqueness.
\vskip 0.3in

\underline{Acknowledgement}Work supported by the NSF of China under Grant No. 19805004 and partially by the NSF of China through C.N. Yang and the Grant LWTZ-1298 of Chinese Academy of Science. The authors have also benefited a lot from the {\it Workshop 2000 on QG-NCYM-String} sponsored by the Center for Advanced Studies of Tsinghua University. The authors acknowledge Dr. J.X. Lu for reading the manuscript.

\end{document}